\documentclass[pre,preprint]{revtex4}
\usepackage{amssymb}
\usepackage{graphicx}
\usepackage{hyperref}
\usepackage{bm}

\def\x{{\bm x}}

\def\hp{\hat\phi}
\def\tf{\tilde f}
\def\tF{\tilde F}

\def\sech{\,{\rm sech}}
\def\Real{{\rm Re}}\def\Imag{{\rm Im}}
\ifx\affiliation\undefined 
\def\affiliation#1{\date{\normalsize #1\\ \today}}
\usepackage[margin=1in]{geometry}
\usepackage[sort&compress,numbers]{natbib}
\else 
\fi
\begin{document}

\title{Transverse instability magnetic field thresholds of electron
  phase-space holes}

\author{I H Hutchinson}

\affiliation{Plasma Science and Fusion Center, MIT, Cambridge, MA
  02139, USA}

\ifx\altaffiliation\undefined\maketitle\fi 
\begin{abstract}
  A detailed comparison is presented of analytical and
  particle-in-cell (PIC) simulation investigation of the transverse
  instability, in two dimensions, of initially one-dimensional
  electron phase-space hole equilibria. Good quantitative agreement is
  found between the shift-mode analysis and the simulations for the
  magnetic field ($B$) threshold at which the instability becomes
  overstable (time-oscillatory) and for the real and imaginary parts
  of the frequency. The simulation $B$-threshold for full
  stabilization exceeds the predictions of shift-mode analysis by 20
  to 30\%, because the mode becomes substantially narrower in spatial
  extent than a pure shift. This threshold shift is qualitatively
  explained by the kinematic mechanism of instability.
\end{abstract}
\ifx\altaffiliation\undefined\else\maketitle\fi  

\section{Introduction}

An electron phase-space hole in a collisionless plasma is an isolated
positive electric potential structure with trapped electron orbits on
which the phase-space density is lower than for untrapped orbits. The
resulting local decrease of electron charge density self-consistently
sustains the enhanced potential in steady
state\cite{Turikov1984,Schamel1986a,Eliasson2006,Hutchinson2017}. Electron
holes are formed during the non-linear (electron trapping) phase of
most electrostatic (Penrose) instabilities in one
dimension\cite{Hutchinson2017}. They have also been widely observed in
space plasmas, since electric field sampling rates began to be
fast enough to resolve the rapid passage of a hole past the
spacecraft\citep{Matsumoto1994,Ergun1998,Bale1998,Mangeney1999,Pickett2008,Andersson2009,Wilson2010,Malaspina2013,Malaspina2014,Vasko2015,Mozer2016,Hutchinson2018b,Mozer2018}. In
multiple dimensions, electron holes are known from simulations to
break up by what is called the transverse instability: growing
perturbations that vary in the direction transverse to the direction
of particle trapping, and observed in many studies\citep{Mottez1997,Miyake1998a,Goldman1999,Oppenheim1999,Muschietti2000,Oppenheim2001b,Singh2001,Lu2008}. Such
instabilities probably determine the long-term multidimensional
structure of these electrostatic soliton-like objects, and their
ultimate dissipation into the background plasma. Therefore it is vital
to understand the mechanisms involved and their bearing on
experimental observations.

Recent theory of linear kink stability of an initially planar electron
hole identified the kinematic mechanism of transverse instability at
low magnetization, and confirmed it by comparison with
simulation\cite{Hutchinson2018}. Detailed mathematical analysis also
showed why instability is suppressed by a sufficiently strong magnetic
field in the trapping direction ($z$), normal to the
plane\cite{Hutchinson2018a}. However a quantitative discrepancy of
approximately a factor of two was noted between the magnetic field
threshold calculated analytically and what is observed in PIC
simulations. In simulations, full suppression of the kink instability
takes approximately twice as strong a magnetic field as was found
analytically in that work. The purpose of this paper is to report
further investigations that now explain the prior discrepancy
concerning the threshold, confirming and extending our understanding
of the instability mechanisms.

\begin{figure}[htp]
  \includegraphics[width=0.45\hsize]{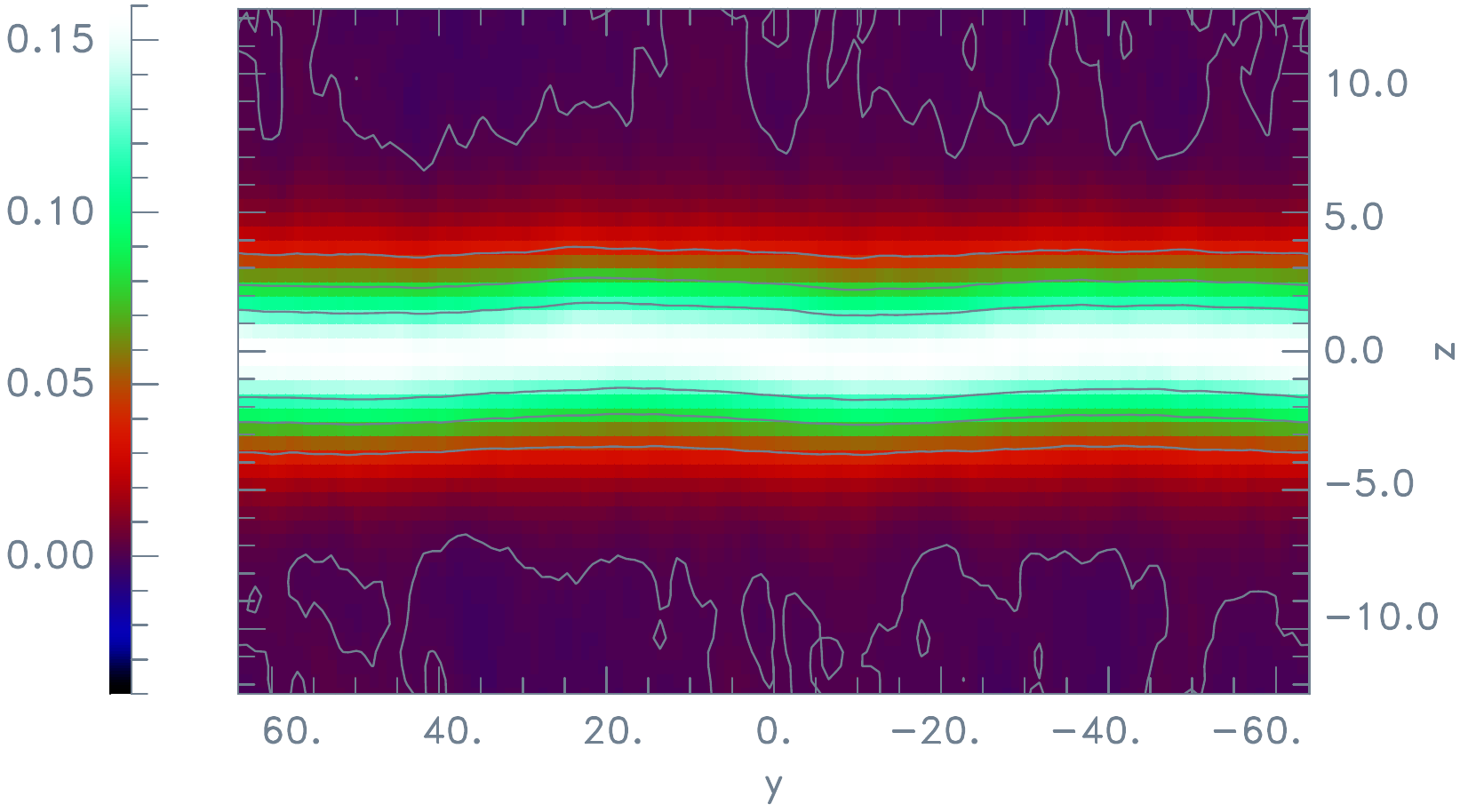}\hskip-0.4\hsize(a) $t=194$\hskip0.3\hsize
  \includegraphics[width=0.45\hsize]{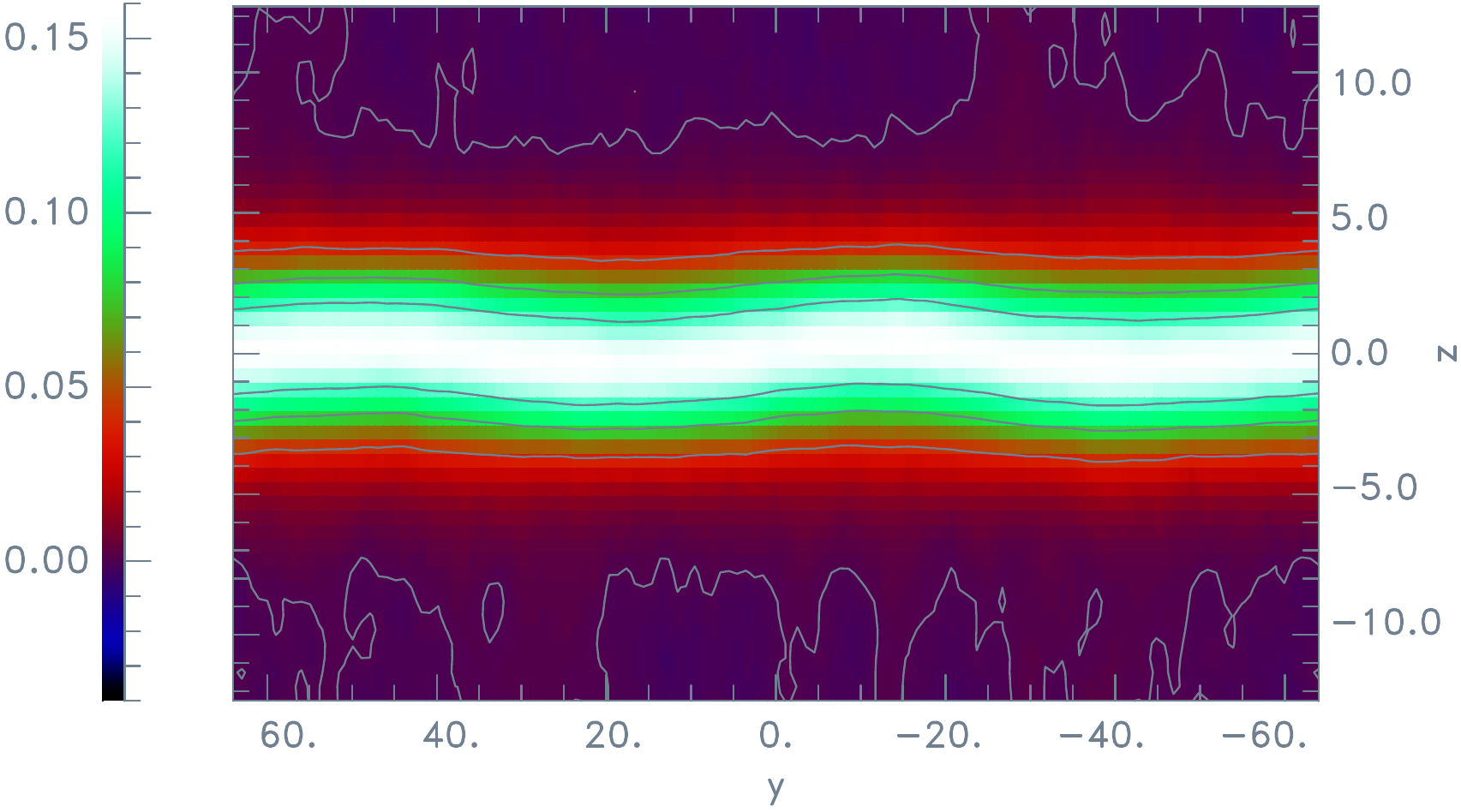}\hskip-0.4\hsize(b) $t=238$\hskip0.3\hsize\ 
  \caption{Example PIC simulation of a kink instability of an
    initially planar hole. Contours of potential $\phi$ in two spatial
    dimensions: $z$ the direction of trapping and the magnetic field,
    $y$ the transverse direction. The perturbation consists of
    sinusoidal wave variation in the $y$ direction of a displacement
    in the $z$-direction. This case has a displacement that is a
    growing standing wave whose amplitude in (b) at later time $t=238$
    is opposite in sign and greater in magnitude than that in (a) at
    $t=194$. ($\psi=0.16$, $\Omega=0.2$.)}
  \label{fig:kinkosc}
\end{figure}
To clarify what is meant by a kink, the spatial form of the
instability is illustrated by the potential contours at two times
during instability growth during a PIC simulation in Fig.\
\ref{fig:kinkosc}. The initial equilibrium is independent of the
transverse coordinate $y$, but as time passes, a growing displacement
of the hole in the $z$ direction develops, which is sinusoidal in $y$,
and in this case oscillating in time.  The two-dimensional simulation
is independent of the third coordinate ($x$) throughout.

Two main factors in the previous assumptions
and approximations are found to cause the discrepancy: (1) It was
previously assumed that solutions to the analytic dispersion relation
are purely growing modes whose frequency $\omega$ is pure imaginary;
it turns out, in contrast, that there are overstable modes at fields a
little above the threshold for stability of the purely growing
modes. And (2) the analytic calculation approximates the eigenmode to
be a pure ``shift-mode'', consisting of a lateral motion independent of
$z$; the PIC simulations show that at the highest unstable magnetic
fields, the eigenmode is not a pure shift; the displacement is
concentrated in the center of the hole.  This effect allows the
overstable modes to exist at somewhat higher fields than for pure
shift-modes.

The present paper draws heavily on the previous analysis, but removes
the restriction to imaginary $\omega$ and gives direct comparisons
with detailed simulations in the vicinity of the threshold.  It does
\emph{not} address or explain the slower-growing instabilities observed
in PIC simulations at high magnetic fields well above the
threshold\cite{Goldman1999,Oppenheim1999,Newman2001a,Oppenheim2001b,Berthomier2002,Lu2008,Wu2010},
which are associated with hole coupling to long-wavelength
perturbations often called streaks or whistlers.

\section{Analytic dispersion relation}

We only summarize here the analysis that was carried out in ref.\
\cite{Hutchinson2018a}, which should be consulted for the mathematical
details.  Ions are taken as a uniform immobile background and only electron
dynamics are included.  Throughout this paper dimensionless units are
used with length normalized to Debye length
$\lambda_D=\sqrt{\epsilon_0T_e/n_0e^2}$, velocity to electron thermal
speeds $v_t=\sqrt{T_e/m_e}$, electric potential to thermal energy
$T_e/e$ and frequency to plasma frequency $\omega_p=v_t/\lambda_D$
(time normalized to $\omega_p^{-1}$).  The linearized analytic
treatment of electrostatic instability of a magnetized electron hole
integrates Vlasov's equation along the equilibrium (zeroth order)
orbits to obtain the first order perturbation to the distribution
function $f_1$ caused by a potential perturbation $\phi_1$. For an
electron hole, the equilibrium is non-uniform in the (trapping)
$z$-direction and differs from the text-book wave case in that only in
the transverse direction ($y$) does Fourier analysis yield uncoupled
eigenmodes. The $z$-dependence of the eigenmodes' potential must be
expressed in a full-wave manner; so we write
\begin{equation}
  \label{eq:eigenmode}
  \phi_1(\x,t)=\hp(z)\exp i(ky-\omega t),  
\end{equation}
taking the transverse wave vector to be in the $y$-direction without
loss of generality.  The integration along unperturbed helical orbits
(the characteristics of the linearized equation) gives rise to an
expansion in harmonics of the cyclotron frequency ($\Omega=eB/m_e$
which represents the magnetic field strength) involving
\begin{equation}
  \label{eq:phim}
  \Phi_m(z,t)\equiv 
  \int_{-\infty}^t \hp(z(\tau)){\rm e}^{-i(m\Omega+\omega)(\tau-t)}d\tau,
\end{equation}
where $z(\tau)=z(t)+\int_t^\tau v_z(t')dt'$ is the position at earlier
time $\tau$. For positive imaginary part of $\omega$
($\omega_i>0$) the parallel distribution function (integrated over
transverse velocities) then can be found as
\begin{eqnarray}\label{eq:f1magnetic}
  f_{\parallel 1}(y,t) =  
  q_e\phi_1(t)\left.{\partial f_{\parallel0}\over\partial W_\parallel}\right|_t
  + \sum_{m=-\infty}^\infty i\left[(m\Omega+\omega)
  {\partial f_{\parallel0}\over \partial W_\parallel}
  +m\Omega {f_{\parallel0}\over T_\perp}\right]
  q_e\Phi_m {\rm e}^{-\zeta_t^2}I_m(\zeta_t^2)
  {\rm e}^{i(ky-\omega t)},
\end{eqnarray}
where $q_e$ and $m_e$ are the electron charge and mass, $W_\parallel$
is the parallel energy ${1\over 2}m_ev_z^2+q_e\phi$, $f_{\parallel0}$
is the unperturbed parallel distribution function, $T_\perp$ is the
perpendicular (Maxwellian) temperature, $\zeta_t$ is the
transverse wavenumber $k$ times the thermal Larmor radius, so that
$\zeta_t^2=k^2T_\perp/\Omega^2m_e$, and $I_m$ is the modified Bessel
function. The first term of this equation is the ``adiabatic''
contribution, which can be thought of as the perturbation that would
have arisen if $f$ had stayed the same function of energy as it was
in the unperturbed equilibrium. The remaining harmonic sum is the
non-adiabatic contribution we denote $\tf_\parallel$.

The specific hole equilibrium we analyze is
\begin{equation}
\label{holeequil}
  \phi_0 = \psi \sech^4(z/4)
\end{equation}
where the constant $\psi$ is the maximum hole potential: the ``depth''
of the hole. The corresponding trapped particle bounce time for the
most deeply trapped electrons is $\omega_b=\sqrt{\psi}/2$. The
background parallel velocity distribution is Maxwellian
of temperature $T_e$, and throughout $T_\perp=T_e$.

The main difficulty is to find the shape of the eigenfunction $\hp(z)$
which self-consistently satisfies the Poisson equation because this is
an integro-differential eigenproblem. For slow time dependence
relative to particle transit time, it can be argued on general grounds
that the eigenmode consists of a spatial shift (by small distance
$\Delta$ independent of position) of the equilibrium potential profile
($\phi_0$) giving:
\begin{equation}
  \label{eq:shiftmode}
  \hp = - \Delta {\partial \phi_0\over \partial z}.
\end{equation}
We must concern ourselves with frequencies whose periods are
not much longer than the transit time; so this shift form cannot be expected to
hold exactly. However, we can obtain a good approximation to the
corresponding eigenvalue of our system by expressing it in terms of a
``Rayleigh Quotient'' which gives an eigenvalue whose errors are only
second order in the deviation of the eigenmode from its exact form.
This mathematial procedure is equivalent to requiring the conservation
of total $z$-momentum under the influence of the assumed shift
eigenmode. The resulting kink of the hole gives rise to two different
contributions to momentum balance: $F_E$ consists of transfer by
Maxwell stress of $z$-momentum in the $y$-direction, and $\tF$
consists of the force exerted by the equilibrium potential on the
non-adiabatic part of the charge density perturbation. The eigenvalue
equation is that they must balance
\begin{equation}
  \label{eq:forcebalance}
  F_E \equiv -\epsilon_0 k^2 \int {d\phi_0\over dz}\phi_1 dz =
 -\int{d\phi_0\over dz}\left( \int q_e\tf_\parallel dv_z\right)   dz
 \equiv   \tF,
\end{equation}
into which we substitute the shift-mode, eqs.\ (\ref{eq:shiftmode} and
\ref{eq:eigenmode}). 

The task then, for specified real $k$, is to find real and imaginary
parts ($\omega_r$ and $\omega_i$) of the complex frequency that
satisfy eqs.\ (\ref{eq:eigenmode}) to (\ref{eq:forcebalance}). If
a solution exists with positive $\omega_i$, it is unstable; if no such
solution exists, the equilibrium is stable. The integrals of eq.\
(\ref{eq:phim}) and (\ref{eq:forcebalance}) are carried out
numerically for given $\omega$. Solutions are displayed
graphically and found precisely by 2-D Newton iteration of $\omega$ to
find the roots of the complex quantity $F_{tot}=\tF-F_E$ when they exist.

\section{Overstability calculated analytically}

\begin{figure}[htp]
  \centering
  {\hsize=0.8\hsize
  (a)\hskip-1.5em\includegraphics[width=0.49\hsize]{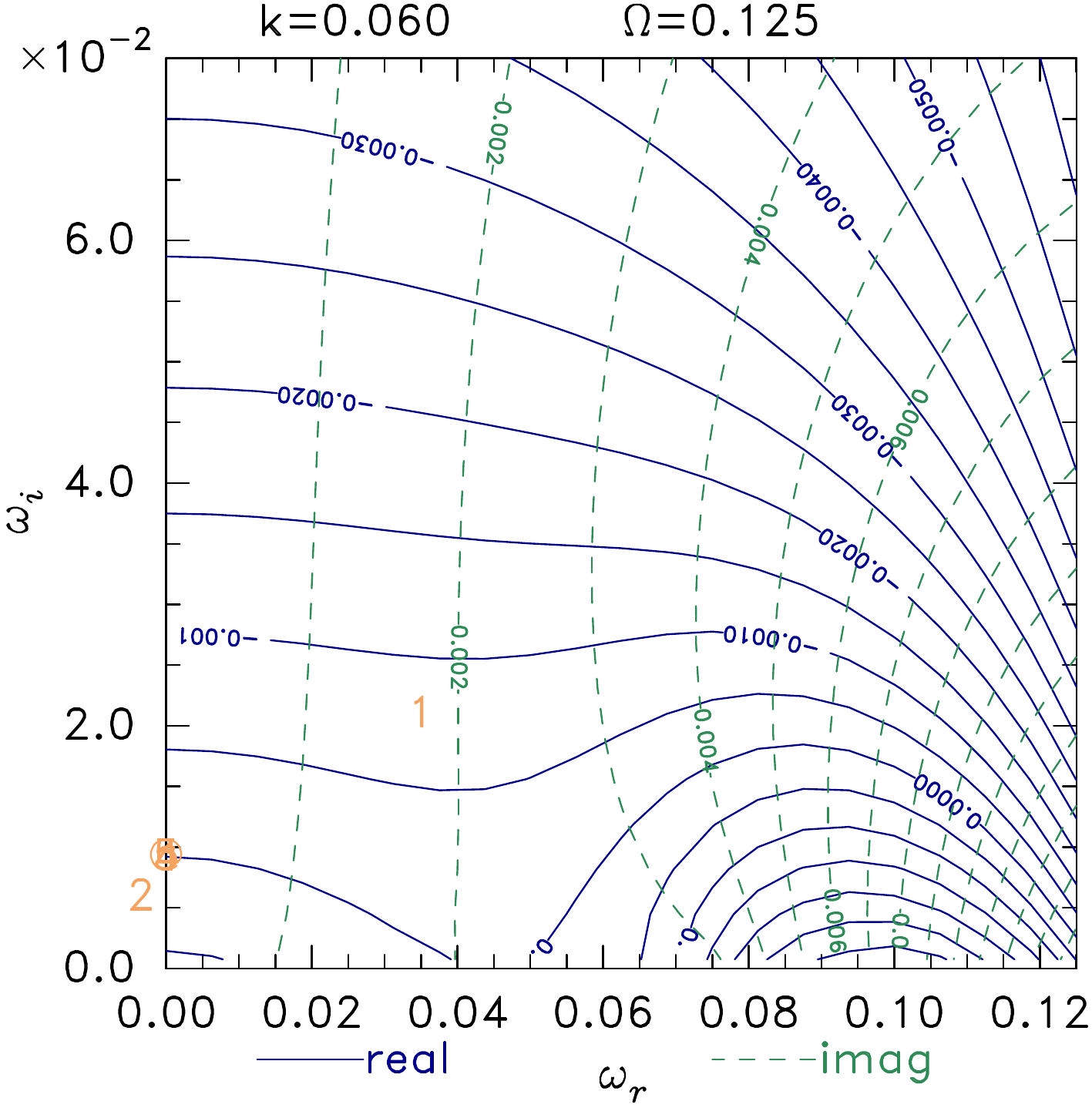}
  (b)\hskip-1.5em\includegraphics[width=0.49\hsize]{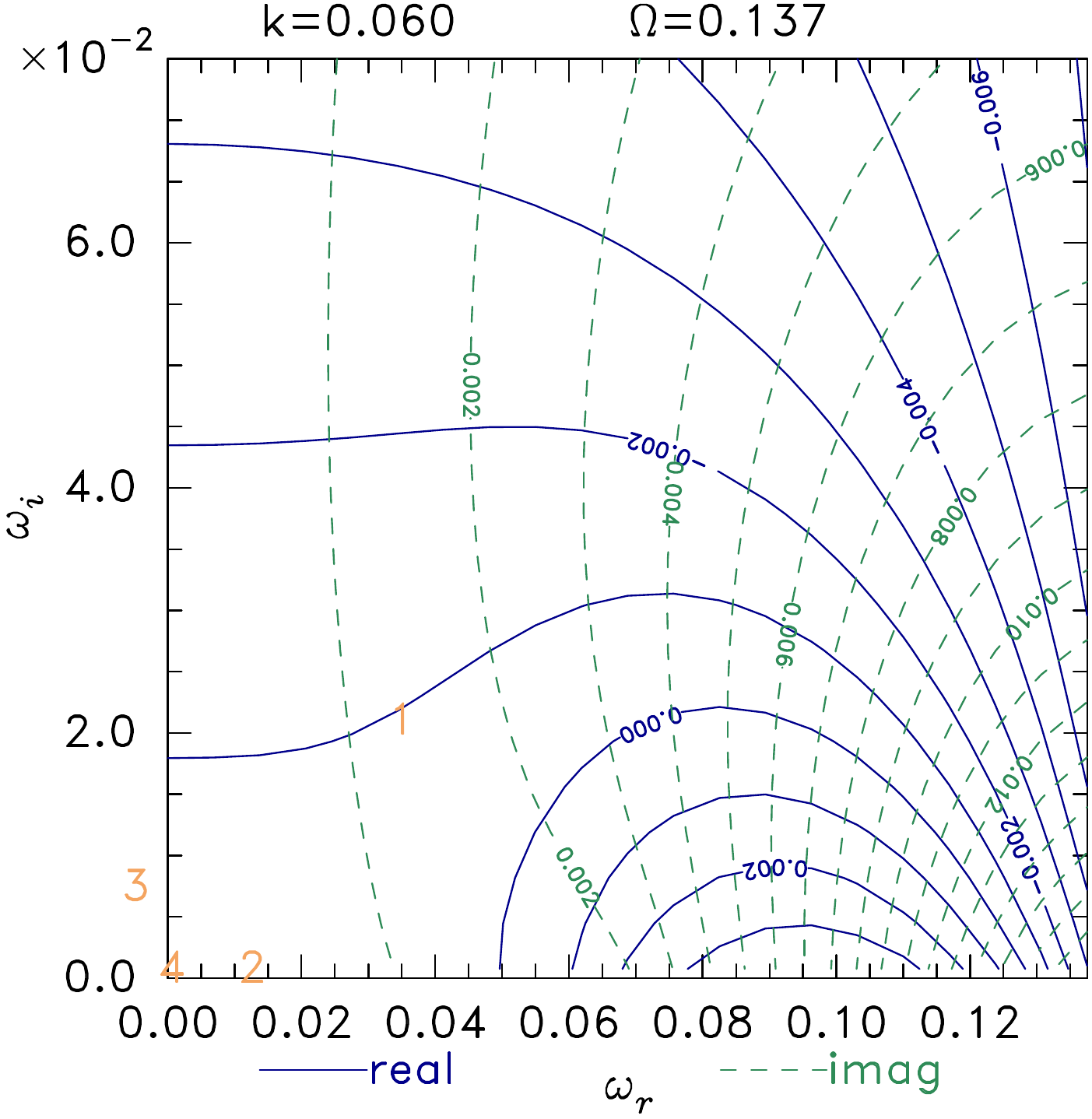}
  (c)\hskip-1.5em\includegraphics[width=0.49\hsize]{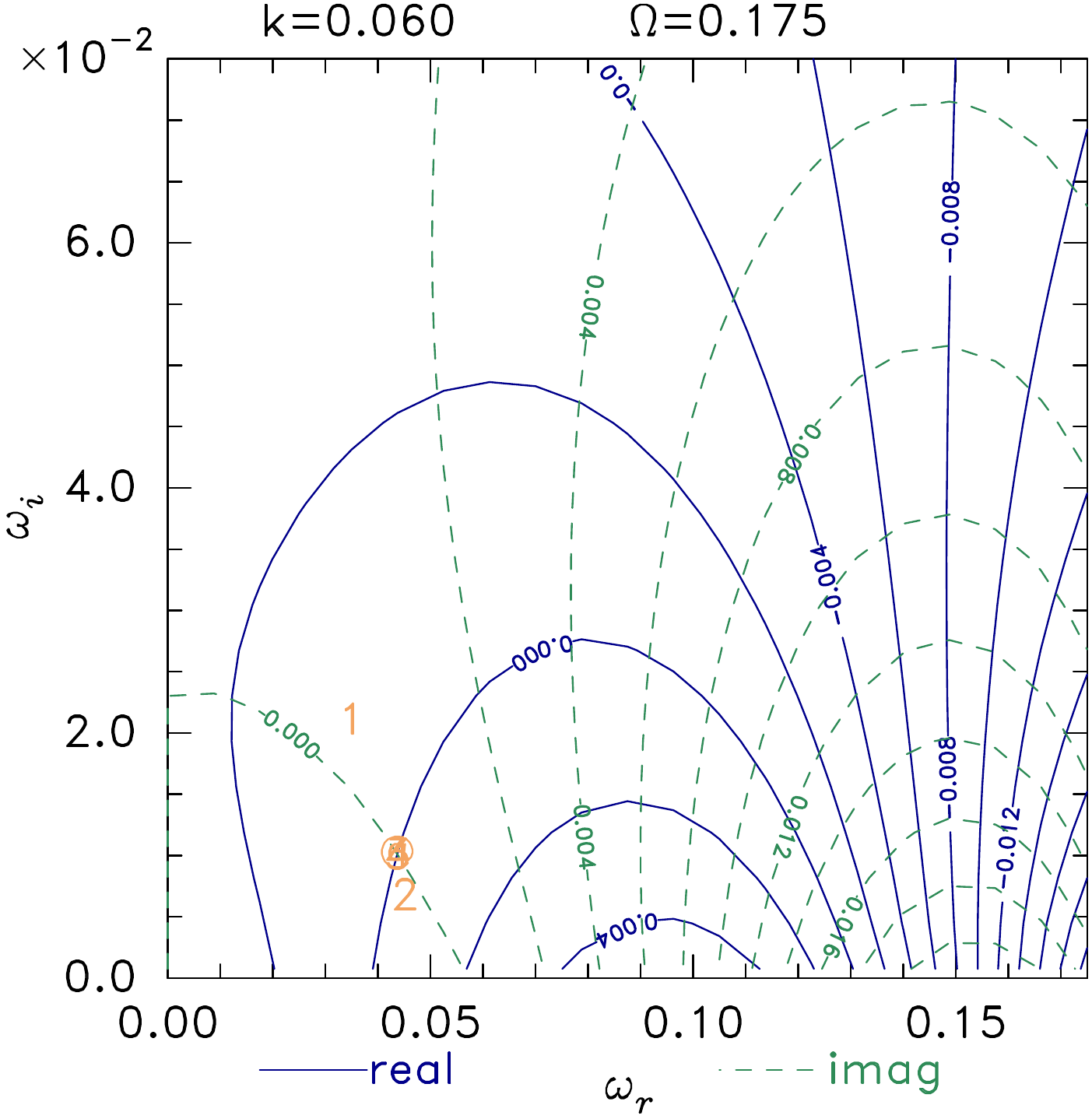}
  (d)\hskip-1.5em\includegraphics[width=0.49\hsize]{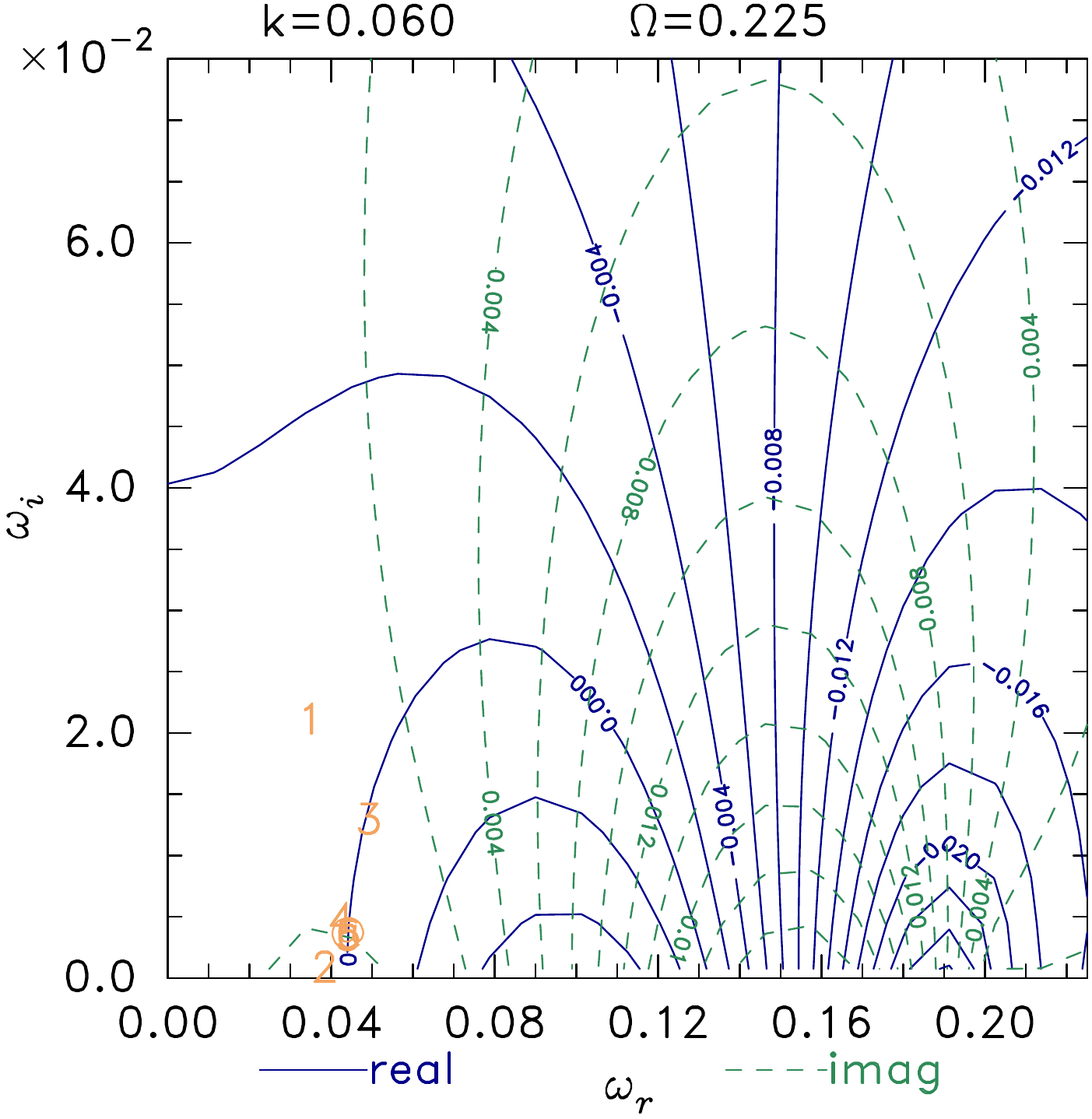}
  }
  \caption{Contours in the complex $\omega$ plane of the real and
    imaginary part of the momentum balance $F_{tot}$, which is
    required to be zero: i.e.\ the solution is the intersection of the
    two zero contours. (a) purely growing case; (b) no solution;
    (c) and (d) overstable solution. Hole parameters $\psi=0.16$,
    $\omega_b=0.2$.}
  \label{fig:contours}
\end{figure}

Figure \ref{fig:contours} shows contours of total (complex) momentum
transfer rate $F_{tot}$ for a shift eigenmode in the complex
$\omega$-plane, for a (real) transverse wave number $k$ chosen to give
a large instability growth rate. It shows where the analytic solutions for
unstable shift kinks occur. 
The solution of the associated dispersion relation, $F_{tot}=0$, is at
the intersection of the two zero contours, for the real and the
imaginary parts. In Fig.\ \ref{fig:contours}(a) and (b) the
$\Imag(F_{tot})=0$ contour exists only at $\omega_r=0$, the imaginary
axis. In (a) a solution exists (with $\omega_i\simeq 0.01$) shown by
the encircled convergence point of the Newton iteration. Convergence
requires only a few iterations; the position of each iterative value
is indicated by the iteration number. In (b), at slightly higher
field, no solution exists because the $\Real(F_{tot})=0$ contour is
absent at $\omega_r=0$. Newton iteration takes the root search into
the negative $\omega_i$ region, where the integration technique is
invalid (and the mode would be damped). This instability suppression
occurs at the previously documented analytic threshold
$\Omega\simeq 0.68\omega_b$, which for this hole depth ($\psi=0.16$)
is $\Omega\simeq 0.136$ ($\omega_p$ units).

However, as $\Omega$ is increased further, for example Fig.\
\ref{fig:contours}(c), a new solution appears away from the imaginary
axis, because the $\Imag(F_{tot})=0$ contour is now present at finite
$\omega_r$. This type of solution persists up to a second threshold at
fields just above the case (d) when the $\Imag(F_{tot})=0$ contour
disappears again below $\omega_i=0$ at the relevant $\omega_r$. The
contour plot domains are deliberately chosen large to illustrate that
there are no other roots. Plots like these also confirm there are no
solutions at higher $\Omega$.

\begin{figure}[htp]
  \centering
  (a)\hskip-1.5em\includegraphics[width=0.48\hsize]{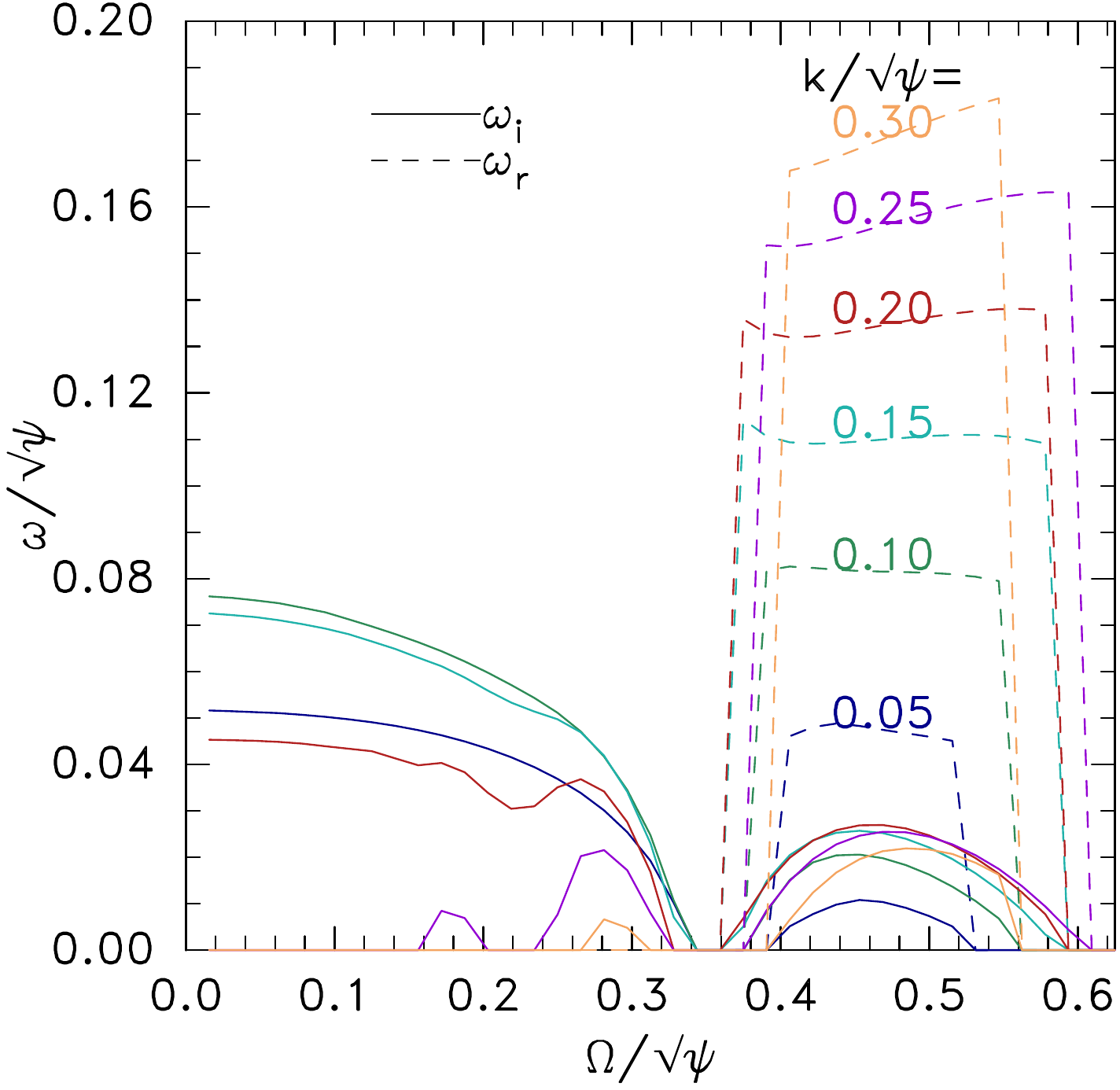}
  (b)\hskip-1.5em\includegraphics[width=0.48\hsize]{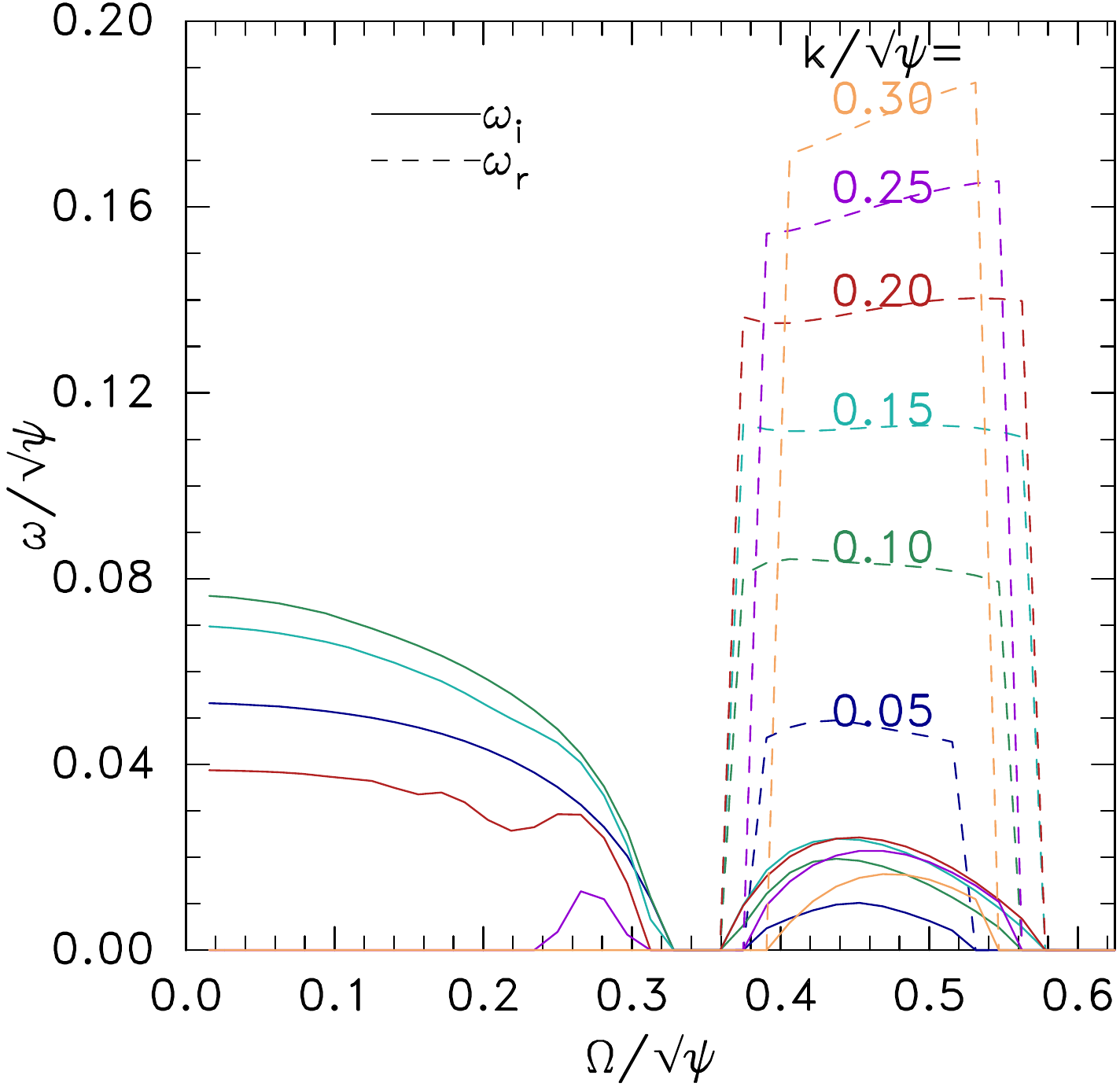}
  \caption{Summary of dispersion solutions for (a) $\psi=0.16$ and (b)
   $\psi=0.36$, and six $k$-values.}
  \label{fig:kvsomegac}
\end{figure}

The dependence of the solutions on the transverse wave number can be
seen by plotting the solution $\omega$ as a function of $\Omega$,
as illustrated in Fig.\ \ref{fig:kvsomegac}. Here we show values
normalized to $\sqrt{\psi}$ because as was noted in prior analysis
this makes the curves almost independent of hole depth (at least for
shallow holes). This near-universality is evident by inspection
of the two cases (a) $\psi=0.16$ and (b) $\psi=0.36$.
The purely growing mode, which exists for
$\Omega/\sqrt{\psi}\lesssim0.33$, is just what was analyzed before and
the results are exactly the same. It is unstable from small $k$ to
about twice the $k$ having maximum growth rate:
$k\simeq \sqrt{\psi}/8$. Purely growing modes for all $k$'s stabilize
at $\Omega\simeq 0.33\sqrt{\psi}$. Then overstable modes (undiscovered
by prior analysis) with $\omega_r\sim k$ appear at marginally higher
$\Omega$. It is uncertain, in view of rounding errors and other
numerical limitations especially at high $k$, whether the apparent gap
here between the two types of instability is really present; in any
case it is narrow. The overstable mode's $\omega_r$ is essentially
independent of $\Omega$, but the growth rate has a continuous
profile. The fastest growing oscillatory mode at fixed $\Omega$ has
$k/\sqrt{\psi}$ value of $0.15-0.2$ ($\omega_r\simeq.11-.14$) at the
lower end of the unstable $\Omega$-range, but $0.20-0.25$
($\omega_r=.14-.16$) at the upper end.  The latter, higher-$k$, modes
also remain unstable to higher values of $\Omega$, even when
$k$-values are high enough to suppress the purely growing,
low-$\Omega$, instability entirely.  The fastest growth rate of the
overstable mode is $\omega_i\simeq 0.025\sqrt{\psi}$.

\begin{figure}[htp]
  \centering
  \includegraphics[width=0.5\hsize]{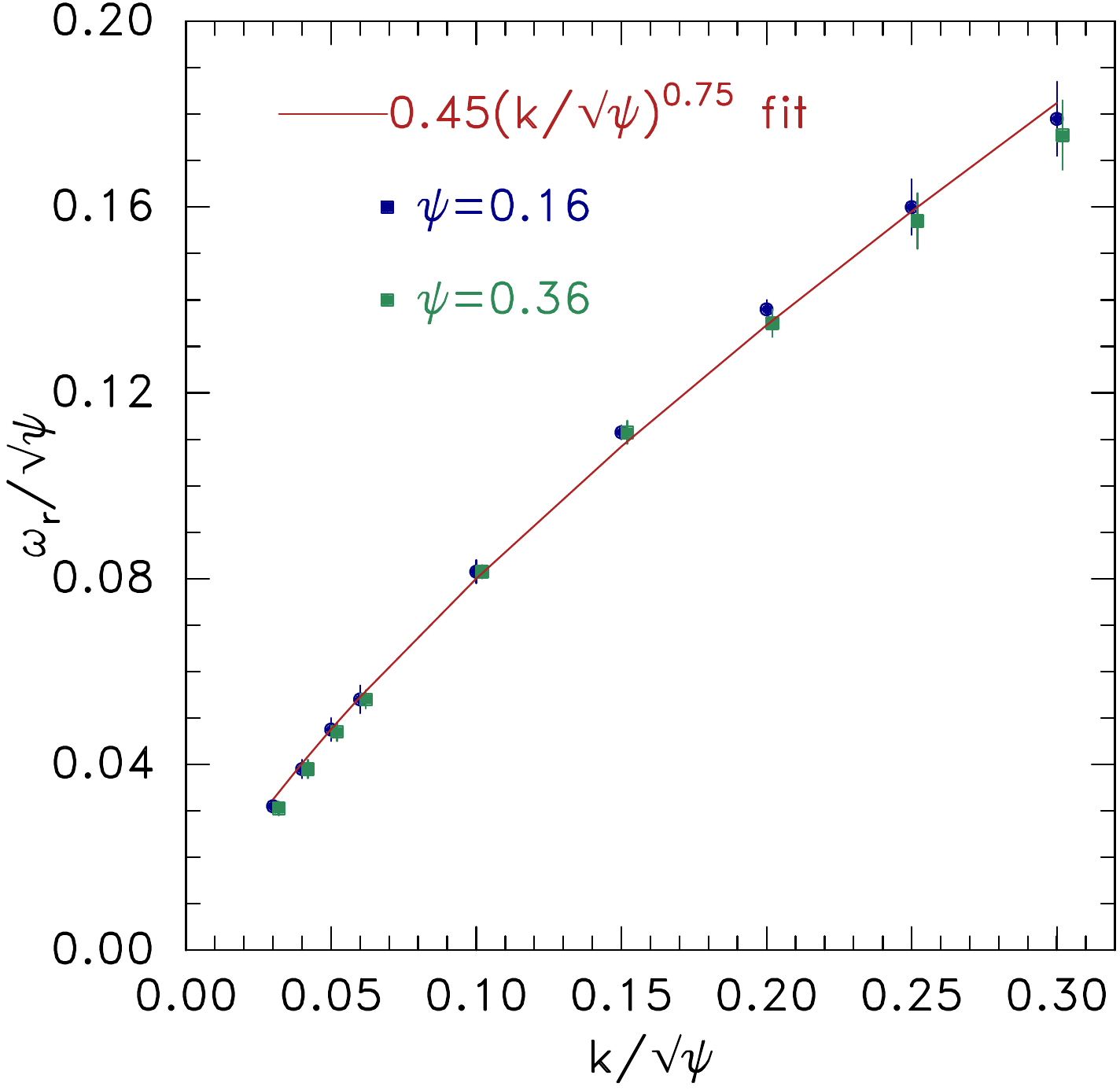}
  \caption{Relation between oscillation frequency $\omega_r$ and
    wavenumber $k$ for overstable modes. Points are derived from
    numerical evaluation of the dispersion relation. The curve is an
    empirical approximate fit to the points.}
  \label{fig:orvkfit}
\end{figure}
The relationship between $k$ and $\omega_r$ can be understood as a
kink vibration of the hole (in the same spirit as was analyzed in
\cite{Newman2001a} for a waterbag hole) that balances $\tF$ and
$F_E$. Reference \cite{Hutchinson2018a} referred to the Maxwell stress
$F_E$ as a ``tension'' force, but that terminology is somewhat
misleading. In fact $F_E$ is a force that acts to \emph{increase} any
kink displacement of the hole, not to \emph{oppose} it. The vibration
then occurs because $\tF$ is such as to \emph{oppose} kinking by
transferring momentum to particles (``jetting'' them) in the direction
of $z$-displacement. Jetting ($\tF$) for low frequency is
approximately proportional to $\omega^2$, that is to acceleration,
while $F_E$ is independent of $\omega$. Therefore their roles have
reversed sign in an analogy with a stretched string. $F_E$ acts like
negative tension (compression) and $\tF$ acts like negative
inertia. The resulting momentum balance is nevertheless a vibrating
wave. A fit to the numerically calculated relationship, approximately
independent of hole depth $\psi$, namely
\begin{equation}\label{eq:fit}
 \omega_r/\sqrt{\psi} \simeq 0.45 (k/\sqrt{\psi})^{0.75}
\end{equation}
is shown in Fig.\ \ref{fig:orvkfit}. The range of $\omega_r$ values is
show by the vertical bars.

In summary, the shift-mode analytic results show there are two
threshold values of magnetic field. The purely growing mode is
suppressed for $\Omega\gtrsim0.34\sqrt{\psi}\simeq0.68\omega_b$. The
overstable mode appears immediately above this threshold and is
present until $\Omega\gtrsim0.6\sqrt{\psi}\simeq1.2\omega_b$.  This
overstable mode, not considered in the previous analysis, explains why
instability and subsequent hole decay is observed in simulations above
the previous pure-growth threshold.

\section{Two-Dimensional PIC simulations}

Particle in cell simulations were carried out with the code
COPTIC\cite{Hutchinson2011a}\footnote{Available from
  https://github.com/ihutch/COPTIC} to compare with the analytic
results. The potential is represented on a periodic two dimensional
domain $-32<z<32$ (trapping direction), $-64<y<64$ (transverse
direction), and initialized with a $\phi_0=\psi\sech^4(z/4)$ hole
uniform in the $y$-direction by seeding the self-consistent particle
distribution calculated by the integral equation method\cite{Hutchinson2017}. Ions are a
uniform immobile background.  In total 0.8 billion electron particles
on 512 parallel cpus are pushed with a time step of 0.5
($\omega_p^{-1}$), and cell size $0.5\times1$ ($\lambda_D$) until
instability growth is observed. All three electron velocity components
are tracked and the particle positions are also periodic.

It is convenient to Fourier transform the resulting potential
perturbations in the $y$-direction, so that
$\phi(y,z,t)=\sum_\ell A_\ell(z,t) {\rm e}^{ik_\ell y}$. Mode 0
($k=0$)
is then the mean, which is essentially the equilibrium (till nonlinear
hole collapse occurs, which we do not investigate here). The higher
modes, $\ell$,
are the possible periodic eigenmodes $k=k_\ell=2\pi
\ell/128$. The modes'
$z$
variation is retained and so each mode depends on time $t$
and position $z$.
Generally a dominant integer mode number is observed to grow, together
with adjacent mode numbers of lower amplitude (and different
frequency).

\begin{figure}[htp]
  \centering
  {\hsize=0.5\hsize
  \includegraphics[width=\hsize]{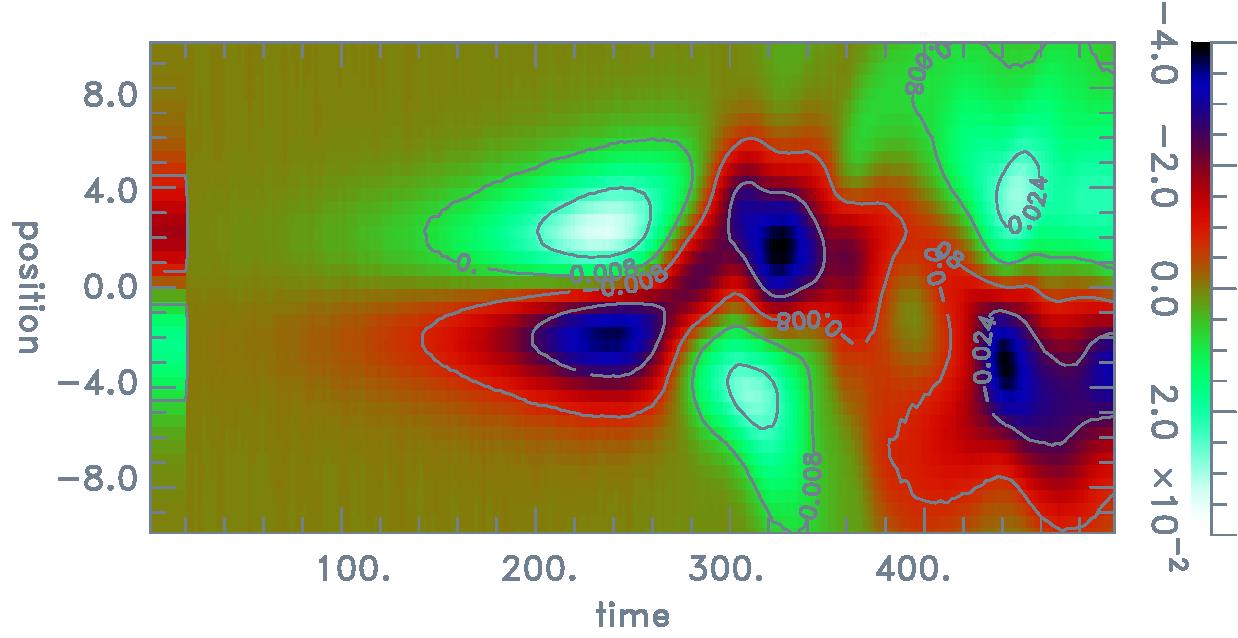}\hskip-\hsize(a)\par
  \includegraphics[width=\hsize]{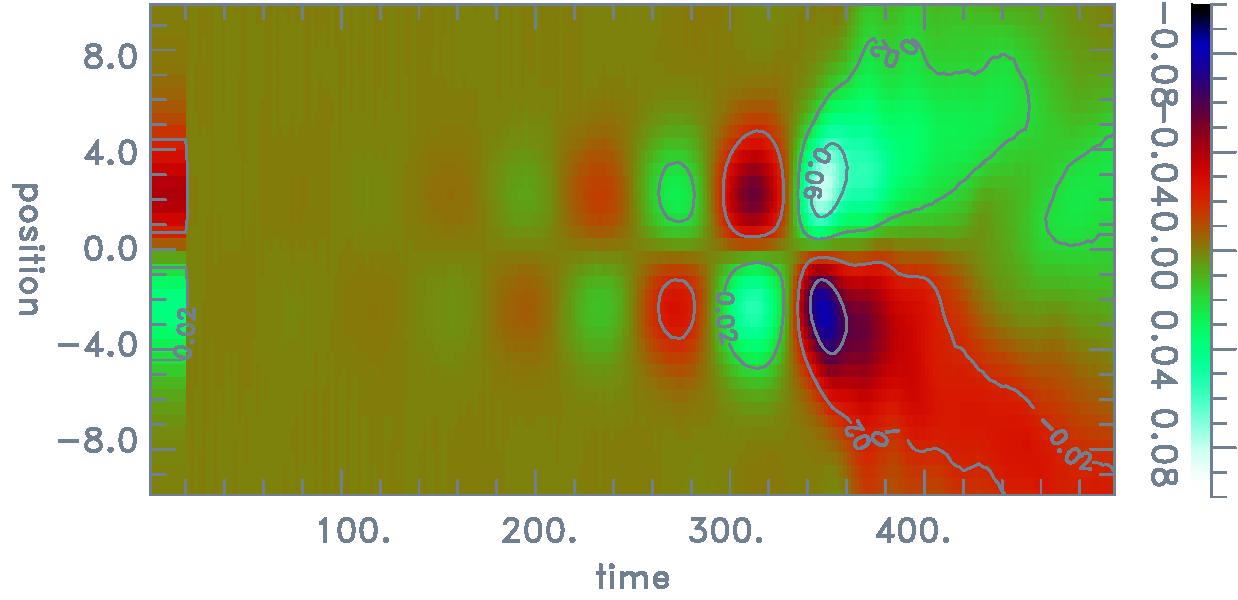}\hskip-\hsize(b)\par
  \includegraphics[width=\hsize]{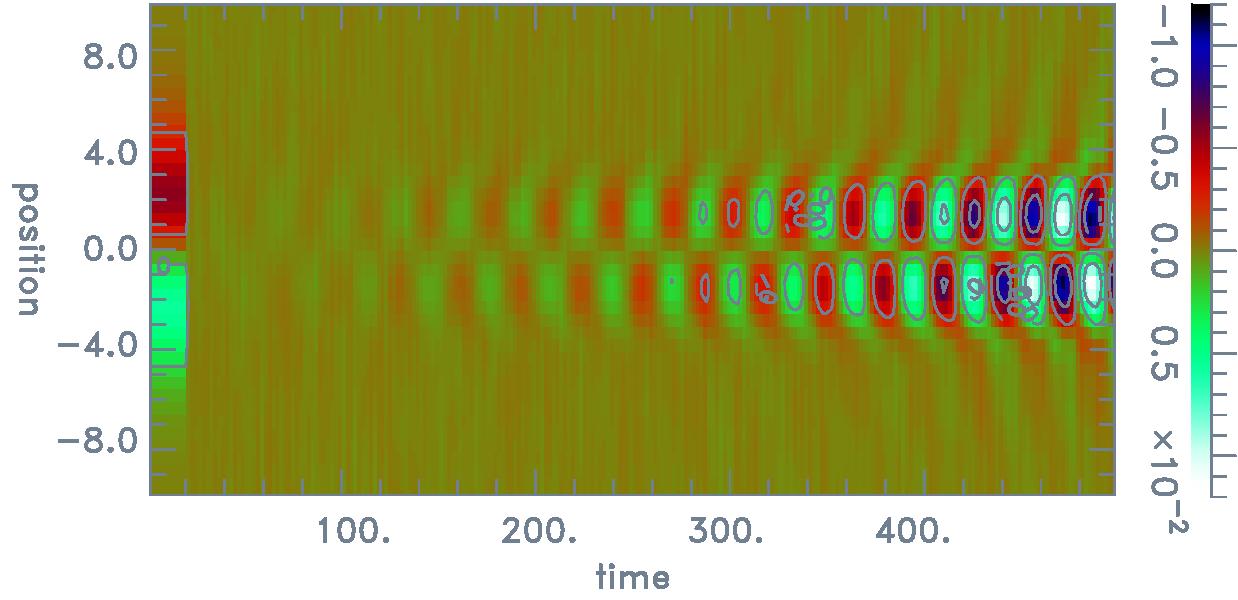}\hskip-\hsize(c)\par
}
\caption{Examples of unstable potential modes in hole simulations with
  $\psi=0.49$. (a) $\Omega=0.2$, $\ell=1$, purely growing. (b)
  $\Omega=0.25$, $\ell=2$, oscillatory. (c) $\Omega=0.45$, $\ell=6$,
  oscillatory. The left-hand section $0<t<20$ is replaced by the shift
  mode $z$-profile for comparison with the PIC results. The color bars
  at right indicate mode amplitudes $A_\ell$.}
  \label{fig:modesunstable}
\end{figure}

Fig.\ \ref{fig:modesunstable} shows example contour plots of the
dominant mode evolution. Only positions $-10<z<10$ close to the hole
are shown. The segment of the plot prior to time 20 (where there is no
significant perturbation yet) has been replaced with a template that
represents an arbitrarily scaled version of the shift-mode
$-{\partial\phi_0\over\partial z}$ to compare with the later observed
structures.
Panel (a) has a magnetic field strength $\Omega=0.20$, just below what
is needed to stabilize the purely growing mode. It grows without
oscillations until nonlinear collapse of the hole sets in at
approxmately $t=250$. Panel (b) has $\Omega=0.25$ in which the purely
growing mode has disappeared and an overstable mode with $\ell=2$ has
replaced it. Both these panels have mode positional structure similar
to the shift-mode. Panel (c) is at $\Omega=0.45$, the uppermost value
of this overstable modes series. It has a much higher mode number:
$\ell=6$. It can also be observed that the $z$-extent of mode (c) is
substantially reduced. Its amplitude decreases to \emph{zero} at a position
that is near the \emph{peak} of the shift-mode template, with only a small
perturbation of reversed polarity outside it. The eigenmode structure
is nearly a factor of 2 narrower in $z$ than the shift-mode template.

\begin{figure}[htp]
  \centering
  \includegraphics[width=0.5\hsize]{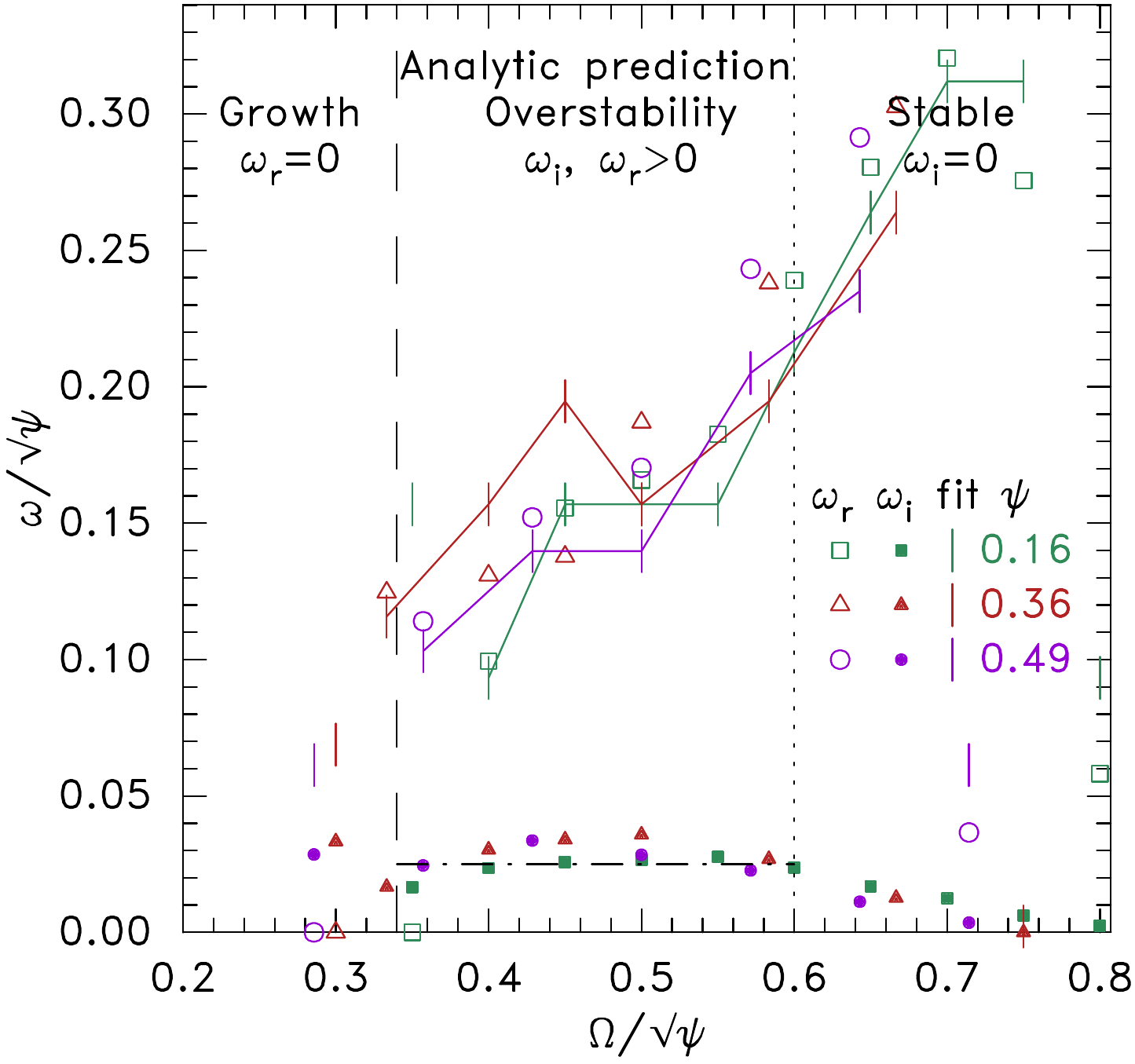}
  \caption{Simulation observations of unstable real (open points) and
    imaginary (filled points) frequencies as a function of magnetic
    field strength. Analytic predictions are shown by dashed lines
    separating the three $\Omega$ regions and showing the growth rate
    of overstability. Vertical bars joined by lines are the analytic frequency
    fit eq.\ (\ref{eq:fit}) for the dominant $k$.}
  \label{fig:ovO}
\end{figure}

We obtain from many simulations' mode results like these the real and
imaginary parts of the mode frequency $\omega_r$ and $\omega_i$, and
the dominant $k$. Fig.\ \ref{fig:ovO} shows the normalized frequencies
as a function of normalized field. 
Analytic prediction of regions are indicated, separated by the
analytic stability thresholds for suppression of the purely growing
mode $\Omega/\sqrt{\psi}=0.34$, and the overstable mode
$\Omega/\sqrt{\psi}=0.6$. The normalization by $\sqrt{\psi}$ brings
the points representing PIC results for different $\psi$ into
approximately universal curves.  The overstable analytic growth rate
$\omega_i/\sqrt{\psi}=0.025$ is shown by the horizontal dash-dot line. It is in
good agreement with the observed PIC growth rates (filled points). The
transition between pure growth and overstability lies between the
lowest two $\Omega$ values for each $\psi$ and is in good agreement
with the analytic prediction (dashed line). However, the overstable
oscillations persist somewhat above the predicted analytic threshold
of stability (short-dashed line). In this right-hand region, the growth
rate falls until it reaches (near) zero when $\Omega$ is 20-25\% past
the 0.6 threshold.

The real frequency $\omega_r$ is shown by open points. The comparison
with the fit (\ref{eq:fit}) to analytic results is provided by bars,
denoting the predicted frequency for the observed dominant $\ell$
value.  The bars expected to agree with PIC observations are joined by
lines, and do indeed fit the observed frequencies well, considering
the discreteness of $k$ enforced by the finite $y$-extent. The
lowest-$\Omega$ (unconnected) bars correspond to zero frequency purely
growing modes, so they do not fit. The highest-$\Omega$ cases show a
sudden jump down to $\ell=1$, and hence much decreased
$\omega_r$. They have extremely low $\omega_i$ and are effectively
stable in respect of the modes treated here
analytically.

\section{Discussion}

In summary, the magnetic-field suppression of the purely growing
instability of electron holes occurs in PIC simulations at a value of
cyclotron frequency $\Omega=0.34\sqrt{\psi}\,\omega_p$ in good
agreement with shift-mode analytic theory. That theory, extended to
account for oscillatory modes, also agrees well with the real and
imaginary parts of observed frequency in the overstable
regime. However, the overstability in simulation persists 20-25\%
beyond the $\Omega\simeq0.6\sqrt{\psi}\,\omega_p$ at which the
shift-mode becomes analytically stable. Spatial narrowing of the
eigenmode compared with the assumed shift-mode appears to be the
reason. Deviation from shift-mode may also explain the fact that no
stable region is observed in simulations between the pure-growth and
overstable regions.

The instability of the high-frequency overstable modes into the region
above the threshold of stability of the shift-mode analytic
calculation can be understood qualitatively as permitted by the
reduction of the $z$-width of the actual unstable modes relative to
the shift-mode shape. In \cite{Hutchinson2018a} it was shown that magnetic
field stabilization of the shift-mode arises because for cyclotron
harmonic $|m|=1$ the contributions from $\Phi_m$ to the trapped
particle momentum change their sign across the resonance
$m\Omega+\omega=\omega_{bounce}$, where $\omega_{bounce}$ is a
function of the trapped particle total energy ($W_\parallel$), that is, of the
particle depth in the potential energy well. For a given value of
$m\Omega+\omega$, particles bouncing slower than this frequency do not
contribute destabilizing trapped momentum. As $m\Omega+\omega$
increases and more trapped particles cease contributing, eventually
the balance between trapped and passing momentum jetting is broken by
this exclusion and the instability is suppressed. However, the exact
value of $\Omega$ at which full suppression occurs depends on the
$z$-extent of the eigenmode as well as on $\omega$. A narrower-$z$
eigenmode can substantially decrease the momentum perturbation
contribution from shallowly trapped and passing electrons, whose
orbits extend past the narrow eigenmode, while decreasing little the
contribution from deeply trapped particles, whose orbits do not. This
effect thus tends to make narrower eigenmodes more unstable than the
shift-mode. So they can persist at magnetic fields beyond the
shift-mode threshold. This sort of consideration also
helps explain qualitatively why oscillatory modes are unstable at
higher $\Omega$ than modes of lower (or zero) $\omega_r$. Instability
is driven by particles and $m$ satisfying
$|m\Omega+\omega|<\omega_{bounce}$. The lowest values of
$|m\Omega+\omega|$ occur for $m=-1$ and \emph{highest} $\omega_r$; so
higher $\omega_r$ is liable to be more unstable because more particles
contribute to unstable trapped momentum jetting.

Although using the Rayleigh Quotient (total momentum conservation)
minimizes the dependence of the eigenvalue on the assumed shape of the
eigenmode, it does not suppress it entirely, especially when the
deviation from the shift-mode is large.  The present comparison shows
that the shift-mode assumption becomes quantitatively inadequate at
high magnetic field. The plasma always finds the most unstable mode,
and the shift-mode is no longer exactly it. The eigenmode structure of
the coupled hole-wave (streak or whistler) instability observed in
simulations at even higher magnetic fields deviates from the
shift-mode assumption in a different way. The shift-mode presumes
there is no net potential difference across the hole, the wave
coupling violates that assumption. It therefore seems likely that the
hole-wave instability requires a more complex eigenmode for its
analysis, probably involving the ingredients described by
\cite{Newman2001a}. A comprehensive analytic understanding of
that instability is as yet lacking.

\section*{Acknowledgements}

Helpful discussions with Xiang Chen and David Malaspina are gratefully
acknowledged.  This work was partially supported by NASA grant
NNX16AG82G.

\bibliography{MyAll}

\begin{thebibliography}{31}
\expandafter\ifx\csname natexlab\endcsname\relax\def\natexlab#1{#1}\fi
\def\au#1{#1} \def\ed#1{#1} \def\yr#1{#1}\def\at#1{#1}\def\jt#1{\textit{#1}}
  \def\bt#1{#1}\def\bvol#1{\textbf{#1}} \def\vol#1{#1} \def\pg#1{#1}
  \def\publ#1{#1}\def\arxiv#1{#1}\def\org#1{#1}\def\st#1{\textit{#1}}

\bibitem[Andersson {\em et~al.\/}(2009)Andersson, Ergun, Tao, Roux, Lecontel,
  Angelopoulos, Bonnell, McFadden, Larson, Eriksson, Johansson, Cully, Newman,
  Goldman, Glassmeier \& Baumjohann]{Andersson2009}
{\sc \au{Andersson, L.}, \au{Ergun, R.~E.}, \au{Tao, J.}, \au{Roux, A.},
  \au{Lecontel, O.}, \au{Angelopoulos, V.}, \au{Bonnell, J.}, \au{McFadden,
  J.~P.}, \au{Larson, D.~E.}, \au{Eriksson, S.}, \au{Johansson, T.}, \au{Cully,
  C.~M.}, \au{Newman, D.~N.}, \au{Goldman, M.~V.}, \au{Glassmeier, K.~H.} \&
  \au{Baumjohann, W.}} \yr{2009}  \at{{New features of electron phase space
  holes observed by the THEMIS mission}}.  \jt{Physical Review Letters}
  \bvol{102}~(22),  \pg{225004}.

\bibitem[Bale {\em et~al.\/}(1998)Bale, Kellogg, Larsen, Lin, Goetz \&
  Lepping]{Bale1998}
{\sc \au{Bale, S.~D.}, \au{Kellogg, P.~J.}, \au{Larsen, D.~E.}, \au{Lin,
  R.~P.}, \au{Goetz, K.} \& \au{Lepping, R.~P.}} \yr{1998}  \at{{Bipolar
  electrostatic structures in the shock transition region: Evidence of electron
  phase space holes}}.  \jt{Geophysical Research Letters}  \bvol{25}~(15),
  \pg{2929--2932}.

\bibitem[Berthomier {\em et~al.\/}(2002)Berthomier, Muschietti, Bonnell, Roth
  \& Carlson]{Berthomier2002}
{\sc \au{Berthomier, M.}, \au{Muschietti, L.}, \au{Bonnell, J.~W.}, \au{Roth,
  I.} \& \au{Carlson, C.~W.}} \yr{2002}  \at{{Interaction between electrostatic
  whistlers and electron holes in the auroral region}}.  \jt{Journal of
  Geophysical Research: Space Physics}  \bvol{107}~(A12),  \pg{1--11}.

\bibitem[Eliasson \& Shukla(2006)]{Eliasson2006}
{\sc \au{Eliasson, B.} \& \au{Shukla, P.~K.}} \yr{2006}  \at{{Formation and
  dynamics of coherent structures involving phase-space vortices in plasmas}}.
  \jt{Physics Reports}  \bvol{422}~(6),  \pg{225--290}.

\bibitem[Ergun {\em et~al.\/}(1998)Ergun, Carlson, McFadden, Mozer, Muschietti,
  Roth \& Strangeway]{Ergun1998}
{\sc \au{Ergun, R.~E.}, \au{Carlson, C.~W.}, \au{McFadden, J.~P.}, \au{Mozer,
  F.~S.}, \au{Muschietti, L.}, \au{Roth, I.} \& \au{Strangeway, R.~J.}}
  \yr{1998}  \at{{Debye-Scale Plasma Structures Associated with
  Magnetic-Field-Aligned Electric Fields}}.  \jt{Physical Review Letters}
  \bvol{81}~(4),  \pg{826--829}.

\bibitem[Goldman {\em et~al.\/}(1999)Goldman, Oppenheim \& Newman]{Goldman1999}
{\sc \au{Goldman, M.~V.}, \au{Oppenheim, M.~M.} \& \au{Newman, D.~L.}}
  \yr{1999}  \at{{Nonlinear two-stream instabilities as an explanation for
  auroral bipolar wave structures}}.  \jt{Geophysical Research Letters}
  \bvol{26}~(13),  \pg{1821--1824}.

\bibitem[Hutchinson(2011)]{Hutchinson2011a}
{\sc \au{Hutchinson, I.~H.}} \yr{2011}  \at{{Nonlinear collisionless plasma
  wakes of small particles}}.  \jt{Phys. Plasmas}  \bvol{18},  \pg{32111}.

\bibitem[Hutchinson(2017)]{Hutchinson2017}
{\sc \au{Hutchinson, I.~H.}} \yr{2017}  \at{{Electron holes in phase space:
  What they are and why they matter}}.  \jt{Physics of Plasmas}  \bvol{24}~(5),
   \pg{055601}.

\bibitem[Hutchinson(2018{\natexlab{{\em a\/}}})]{Hutchinson2018}
{\sc \au{Hutchinson, I.~H.}} \yr{2018{\natexlab{{\em a\/}}}}  \at{{Kinematic
  Mechanism of Plasma Electron Hole Transverse Instability}}.  \jt{Physical
  Review Letters}  \bvol{120}~(20),  \pg{205101}.

\bibitem[Hutchinson(2018{\natexlab{{\em b\/}}})]{Hutchinson2018a}
{\sc \au{Hutchinson, I.~H.}} \yr{2018{\natexlab{{\em b\/}}}}  \at{{Transverse
  instability of electron phase-space holes in multi-dimensional Maxwellian
  plasmas}}.  \jt{Journal of Plasma Physics}  \bvol{84},  \pg{905840411},
  \arxiv{arXiv: 1804.08594}.

\bibitem[Hutchinson \& Malaspina(2018)]{Hutchinson2018b}
{\sc \au{Hutchinson, I.~H.} \& \au{Malaspina, D.~M.}} \yr{2018}
  \at{{Prediction and Observation of Electron Instabilities and Phase Space
  Holes Concentrated in the Lunar Plasma Wake}}.  \jt{Geophysical Research
  Letters}  \pg{pp. 3838--3845}.

\bibitem[Lu {\em et~al.\/}(2008)Lu, Lembege, Tao \& Wang]{Lu2008}
{\sc \au{Lu, Q.~M.}, \au{Lembege, B.}, \au{Tao, J.~B.} \& \au{Wang, S.}}
  \yr{2008}  \at{{Perpendicular electric field in two-dimensional electron
  phase-holes: A parameter study}}.  \jt{Journal of Geophysical Research}
  \bvol{113}~(A11),  \pg{A11219}.

\bibitem[Malaspina {\em et~al.\/}(2014)Malaspina, Andersson, Ergun, Wygant,
  Bonnell, Kletzing, Reeves, Skoug \& Larsen]{Malaspina2014}
{\sc \au{Malaspina, D.~M.}, \au{Andersson, L.}, \au{Ergun, R.~E.}, \au{Wygant,
  J.~R.}, \au{Bonnell, J.~W.}, \au{Kletzing, C.}, \au{Reeves, G.~D.},
  \au{Skoug, R.~M.} \& \au{Larsen, B.~A.}} \yr{2014}  \at{{Nonlinear electric
  field structures in the inner magnetosphere}}.  \jt{Geophysical Research
  Letters}  \bvol{41},  \pg{5693--5701}.

\bibitem[Malaspina {\em et~al.\/}(2013)Malaspina, Newman, Willson, Goetz,
  Kellogg \& Kerstin]{Malaspina2013}
{\sc \au{Malaspina, D.~M.}, \au{Newman, D.~L.}, \au{Willson, L.~B.}, \au{Goetz,
  K.}, \au{Kellogg, P.~J.} \& \au{Kerstin, K.}} \yr{2013}  \at{{Electrostatic
  solitary waves in the solar wind: Evidence for instability at solar wind
  current sheets}}.  \jt{Journal of Geophysical Research: Space Physics}
  \bvol{118}~(2),  \pg{591--599}.

\bibitem[Mangeney {\em et~al.\/}(1999)Mangeney, Salem, Lacombe, Bougeret,
  Perche, Manning, Kellogg, Goetz, Monson \& Bosqued]{Mangeney1999}
{\sc \au{Mangeney, A.}, \au{Salem, C.}, \au{Lacombe, C.}, \au{Bougeret, J.-L.},
  \au{Perche, C.}, \au{Manning, R.}, \au{Kellogg, P.~J.}, \au{Goetz, K.},
  \au{Monson, S.~J.} \& \au{Bosqued, J.-M.}} \yr{1999}  \at{{WIND observations
  of coherent electrostatic waves in the solar wind}}.  \jt{Annales
  Geophysicae}  \bvol{17}~(3),  \pg{307--320}.

\bibitem[Matsumoto {\em et~al.\/}(1994)Matsumoto, Kojima, Miyatake, Omura,
  Okada, Nagano \& Tsutsui]{Matsumoto1994}
{\sc \au{Matsumoto, H.}, \au{Kojima, H.}, \au{Miyatake, T.}, \au{Omura, Y.},
  \au{Okada, M.}, \au{Nagano, I.} \& \au{Tsutsui, M.}} \yr{1994}
  \at{{Electrostatic solitary waves (ESW) in the magnetotail: BEN wave forms
  observed by GEOTAIL}}.  \jt{Geophysical Research Letters}  \bvol{21}~(25),
  \pg{2915--2918}.

\bibitem[Miyake {\em et~al.\/}(1998)Miyake, Omura, Matsumoto \&
  Kojima]{Miyake1998a}
{\sc \au{Miyake, T.}, \au{Omura, Y.}, \au{Matsumoto, H.} \& \au{Kojima, H.}}
  \yr{1998}  \at{{Two-dimensional computer simulations of electrostatic
  solitary waves observed by Geotail spacecraft}}.  \jt{Journal of Geophysical
  Research}  \bvol{103}~(A6),  \pg{11841}.

\bibitem[Mottez {\em et~al.\/}(1997)Mottez, Perraut, Roux \&
  Louarn]{Mottez1997}
{\sc \au{Mottez, F.}, \au{Perraut, S.}, \au{Roux, A.} \& \au{Louarn, P.}}
  \yr{1997}  \at{{Coherent structures in the magnetotail triggered by
  counterstreaming electron beams}}.  \jt{Journal of Geophysical Research}
  \bvol{102}~(A6),  \pg{11399}.

\bibitem[Mozer {\em et~al.\/}(2016)Mozer, Agapitov, Artemyev, Burch, Ergun,
  Giles, Mourenas, Torbert, Phan \& Vasko]{Mozer2016}
{\sc \au{Mozer, F.~S.}, \au{Agapitov, O.~A.}, \au{Artemyev, A.}, \au{Burch,
  J.~L.}, \au{Ergun, R.~E.}, \au{Giles, B.~L.}, \au{Mourenas, D.}, \au{Torbert,
  R.~B.}, \au{Phan, T.~D.} \& \au{Vasko, I.}} \yr{2016}  \at{{Magnetospheric
  Multiscale Satellite Observations of Parallel Electron Acceleration in
  Magnetic Field Reconnection by Fermi Reflection from Time Domain
  Structures}}.  \jt{Physical Review Letters}  \bvol{116}~(14),  \pg{4--8},
  \arxiv{arXiv: arXiv:1011.1669v3}.

\bibitem[Mozer {\em et~al.\/}(2018)Mozer, Agapitov, Giles \& Vasko]{Mozer2018}
{\sc \au{Mozer, F.~S.}, \au{Agapitov, O.~V.}, \au{Giles, B.} \& \au{Vasko, I.}}
  \yr{2018}  \at{{Direct Observation of Electron Distributions inside
  Millisecond Duration Electron Holes}}.  \jt{Physical Review Letters}
  \bvol{121}~(13),  \pg{135102}.

\bibitem[Muschietti {\em et~al.\/}(2000)Muschietti, Roth, Carlson \&
  Ergun]{Muschietti2000}
{\sc \au{Muschietti, L.}, \au{Roth, I.}, \au{Carlson, C.~W.} \& \au{Ergun,
  R.~E.}} \yr{2000}  \at{{Transverse instability of magnetized electron
  holes}}.  \jt{Physical Review Letters}  \bvol{85}~(1),  \pg{94--97}.

\bibitem[Newman {\em et~al.\/}(2001)Newman, Goldman, Spector \&
  Perez]{Newman2001a}
{\sc \au{Newman, D.~L.}, \au{Goldman, M.~V.}, \au{Spector, M.} \& \au{Perez,
  F.}} \yr{2001}  \at{{Dynamics and instability of electron phase-space
  tubes}}.  \jt{Physical Review Letters}  \bvol{86}~(7),  \pg{1239--1242}.

\bibitem[Oppenheim {\em et~al.\/}(1999)Oppenheim, Newman \&
  Goldman]{Oppenheim1999}
{\sc \au{Oppenheim, M.}, \au{Newman, D.~L.} \& \au{Goldman, M.~V.}} \yr{1999}
  \at{{Evolution of Electron Phase-Space Holes in a 2D Magnetized Plasma}}.
  \jt{Physical Review Letters}  \bvol{83}~(12),  \pg{2344--2347}.

\bibitem[Oppenheim {\em et~al.\/}(2001)Oppenheim, Vetoulis, Newman \&
  Goldman]{Oppenheim2001b}
{\sc \au{Oppenheim, M.~M.}, \au{Vetoulis, G.}, \au{Newman, D.~L.} \&
  \au{Goldman, M.~V.}} \yr{2001}  \at{{Evolution of electron phase-space holes
  in 3D}}.  \jt{Geophysical Research Letters}  \bvol{28}~(9),  \pg{1891--1894}.

\bibitem[Pickett {\em et~al.\/}(2008)Pickett, Chen, Mutel, Christopher,
  Santolı´k, Lakhina, Singh, Reddy, Gurnett, Tsurutani, Lucek \&
  Lavraud]{Pickett2008}
{\sc \au{Pickett, J.~S.}, \au{Chen, L.-J.}, \au{Mutel, R.~L.}, \au{Christopher,
  I.~W.}, \au{Santolı´k, O.}, \au{Lakhina, G.~S.}, \au{Singh, S.~V.},
  \au{Reddy, R.~V.}, \au{Gurnett, D.~A.}, \au{Tsurutani, B.~T.}, \au{Lucek, E.}
  \& \au{Lavraud, B.}} \yr{2008}  \at{{Furthering our understanding of
  electrostatic solitary waves through Cluster multispacecraft observations and
  theory}}.  \jt{Advances in Space Research}  \bvol{41}~(10),  \pg{1666--1676}.

\bibitem[Schamel(1986)]{Schamel1986a}
{\sc \au{Schamel, H.}} \yr{1986}  \at{{Electrostatic Phase Space Structures in
  Theory and Experiment}}.  \jt{Physics Reports}  \bvol{3}~(3),  \pg{161--191}.

\bibitem[Singh {\em et~al.\/}(2001)Singh, Loo \& Wells]{Singh2001}
{\sc \au{Singh, N.}, \au{Loo, S.~M.} \& \au{Wells, B.~E.}} \yr{2001}
  \at{{Electron hole structure and its stability depending on plasma
  magnetization}}.  \jt{Journal of Geophysical Research}  \bvol{106}~(A10),
  \pg{21183--21198}.

\bibitem[Turikov(1984)]{Turikov1984}
{\sc \au{Turikov, V.~A.}} \yr{1984}  \at{{Electron Phase Space Holes as
  Localized BGK Solutions}}.  \jt{Physica Scripta}  \bvol{30}~(1),
  \pg{73--77}.

\bibitem[Vasko {\em et~al.\/}(2015)Vasko, Agapitov, Mozer, Artemyev \&
  Jovanovic]{Vasko2015}
{\sc \au{Vasko, I.~Y.}, \au{Agapitov, O.~V.}, \au{Mozer, F.}, \au{Artemyev,
  A.~V.} \& \au{Jovanovic, D.}} \yr{2015}  \at{{Magnetic field depression
  within electron holes}}.  \jt{Geophysical Research Letters}  \bvol{42}~(7),
  \pg{2123--2129}.

\bibitem[Wilson {\em et~al.\/}(2010)Wilson, Cattell, Kellogg, Goetz, Kersten,
  Kasper, Szabo \& Wilber]{Wilson2010}
{\sc \au{Wilson, L.~B.}, \au{Cattell, C.~A.}, \au{Kellogg, P.~J.}, \au{Goetz,
  K.}, \au{Kersten, K.}, \au{Kasper, J.~C.}, \au{Szabo, A.} \& \au{Wilber, M.}}
  \yr{2010}  \at{{Large-amplitude electrostatic waves observed at a
  supercritical interplanetary shock}}.  \jt{Journal of Geophysical Research:
  Space Physics}  \bvol{115}~(12),  \pg{A12104}.

\bibitem[Wu {\em et~al.\/}(2010)Wu, Lu, Huang \& Wang]{Wu2010}
{\sc \au{Wu, M.}, \au{Lu, Q.}, \au{Huang, C.} \& \au{Wang, S.}} \yr{2010}
  \at{{Transverse instability and perpendicular electric field in
  two-dimensional electron phase-space holes}}.  \jt{Journal of Geophysical
  Research: Space Physics}  \bvol{115}~(10),  \pg{A10245}.

\end{thebibliography}
\end{document}